\documentclass[twocolumn,aps,showpacs,groupedaddress]{revtex4}
\usepackage{array}
\usepackage{graphicx}

\def\be{\begin{equation}}
\def\ee{\end{equation}}
\def\bea{\begin{eqnarray}}
\def\eea{\end{eqnarray}}
\def\ba{\begin{array}}
\def\ea{\end{array}}
\def\bdm{\begin{displaymath}}
\def\edm{\end{displaymath}}

\begin{document}

\title{Phase diagrams of a p-wave superconductor inside a mesoscopic disc-shaped sample}

\author{Bor-Luen Huang and S.-K. Yip}

\affiliation{Institute of Physics, Academia Sinica, Taipei, Taiwan}

\date{\today }

\begin{abstract}
We study the finite-size and boundary effects on a time-reversal-symmetry breaking p-wave
superconducting state in a mesoscopic disc geometry using Ginzburg-Landau theory.
We show that, for a large parameter range,
the system exhibits multiple phase transitions.
The superconducting transition from the normal state can also
be reentrant as a function of external magnetic field.

\end{abstract}


\maketitle

Studies of Fermi superfluids and superconductors with multi-component order parameters
have drawn much attention over the last few decades.  A well-established example is
spin-triplet p-wave superfluid $^3$He \cite{He3}, the order parameter of which is a
complex $3\times3$ matrix instead of a single component scalar as in the case
of s-wave superconductors.   Many unusual behaviors of this superfluid are known,
in particular "textures",
where the order parameter varies in space in a non-trivial way due to external
fields, flows, or confining walls.   There have also been intense
studies of superconductors that are believed also to possess multi-component order
parameters, for example, UPt$_3$ \cite{UPt3} and Sr$_{2}$RuO$_{4}$ \cite{SrRuO}.
While the precise order parameters in both cases are still controversial,
many believe that both these two superconductor have order parameters
which break time-reversal symmetry.  Broken time reversal symmetry
necessarily requires a multi-component order parameter.  A single component
order parameter, though it can belong to a non-trivial one-dimensional
representation, cannot break time-reversal symmetry in  the sense that
its complex conjugate differs from the original one only via a gauge transformation.
A superconductor with broken time reversal symmetry can have exotic properties,
such as circular dichroism and birefringence (Kerr rotation) \cite{Kapitulnik09},
internal magnetic fields \cite{Muzikar} and surface currents \cite{surfacecurrent}, to name a few.
Experiments claimed to support broken time reversal symmetry have been reported
both for UPt$_3$ \cite{Luke93,Strand} and Sr$_{2}$RuO$_{4}$ \cite{Luke98,Kidwingira}, though
negative results are also in the literature \cite{Kambara96,Hicks10}.
Other aspects of intense recent interest are half-quantum vortices \cite{Budakian11}
arising from the spin degrees of freedom,
and Majorana vortex bound states \cite{M}, which would be possible if the order parameter
is a p-wave with the form $p_x+ip_y$, which is the case proposed for Sr$_{2}$RuO$_{4}$.

In this paper, we study a two-component superconductor in a confined geometry.
Our motivations are several folded.  First, as mentioned, in the context of $^3$He, a
confining geometry can induce a non-trivial texture.  A surface is necessarily
a strong breaker of rotational invariance, and hence its effect on the order
parameter depends on the relative orientation between the two.  The energetically
most favorable configuration therefore does {\em not} necessarily correspond
to simply taking the uniform bulk order parameter and suppressing its magnitude
near the surface.  Second, for
a multi-component order parameter, or more precisely
an order parameter that belongs to a multi-dimensional representation,
the different components possess the same transition temperature $T_c^0$ in the
bulk in the absence of external perturbations (by definition).  However,
external perturbations can split the degeneracies, resulting in multiple
phase transitions.  This has been discussed in the context of thin-films
of $^3$He-B \cite{Kawasaki04}, as well as for UPt$_3$ under the influence of the underlying
anti-ferromagnetic order \cite{UPt3}.
There have also been many recent experimental studies of mesoscopic superconductors,
s \cite{meso-s}, p \cite{Cai}, and d wave \cite{meso-d}, including some interesting
theoretical predications for the latter, e.g. \cite{meso-d-t},
but the physics associated with the multi-component nature
of the order parameter is less explored in the literature.

We shall consider a thin circular disc of radius $R$ lying
in the x-y plane.  Variation of the order parameter
along z, as well as
the  magnetic field generated by the supercurrent of the sample,
will be ignored. We shall study a superconductor with a two-component
order parameter. The two components $\eta_{x,y}$ are supposed to
transform as a vector under rotations within the x-y plane.
We shall study how the order parameter varies
over the disc. For definiteness, $\eta_{x,y}$ are taken to be
the two in-plane components of the orbital part of the
p-wave order parameter, that is, the momentum dependence is $\eta_xp_x+\eta_yp_y$.
However, we expect that many of our findings should be
common to other superconductors with multi-component order parameter.
This point will be discussed again below.
Recently, a group \cite{Huo11} has studied theoretically this same system
using Bogoliubov-deGennes equations.  Their results however differ significantly
from ours.  A comparison will be given later.

We shall employ Ginzburg-Landau (GL) theory, but as we shall argue later,
our conclusions are more general.  The GL free energy density (per unit area)
$\mathcal{F}$ consists of several contributions.  The bulk contribution,
$\mathcal{F}_b$, can be written as
\be
\mathcal{F}_b = \alpha (\vec \eta^* \cdot \vec \eta)
   + \beta_1 (\vec \eta^* \cdot \vec \eta)^2 + \beta_2 |\vec \eta \cdot \vec \eta|^2,
   \label{fb}
\ee
where $\alpha=\alpha'(t-1)$ with $\alpha'>0$, $t\equiv{}T/T_c^0$ is
the ratio of the temperature $T$ relative to the bulk transition temperature $T_c^0$,
$\vec\eta=\eta_x\hat{x}+\eta_y\hat{y}$, which we would often denoted
as $(\eta_x,\eta_y)$.  Stability requires
$\beta_1>0$, $\beta_1>-\beta_2$.  We shall take $\beta_2>0$, so that for
the bulk the equilibrium order parameter below $T_c^0$ has the form
$\vec\eta=\Psi(1,\pm{}i)$, so that it has broken time-reversal  symmetry,
with $|\Psi|=\frac{1}{2}(\frac{\alpha'}{\beta_1})^{1/2}$.

In the presence of gradients, there is an additional contribution to $\mathcal{F}$
given by
\be
\mathcal{F}_g = K_1 (D_j \eta_l) (D_j \eta_l)^* + K_2 (D_j \eta_j) (D_l \eta_l)^*
   + K_3 (D_j \eta_l) (D_l \eta_j)^*,
   \label{fg}
\ee
where $D_j\equiv\partial_j+\frac{2ie}{c}A_j$ is the gauge invariant derivative.
$\vec{A}$ is the vector potential, and the electron charge is $-e$.
Repeated indices $j,l$ are summed over $x,y$ in eq.(\ref{fg}).
In writing down eq.(\ref{fb}) and (\ref{fg}), we have ignored crystal anisotropies
for simplicity, but we do not expect significant qualitative change in our predictions below.
The more general forms can be found in, e.g., \cite{VG,Sigrist}.
Stability requires $K_1>0$, $K_{123}\equiv{}K_1+K_2+K_3>0$.
If we take $\eta_{x,y}$ to represent the two in-plane orbital components of
a pure p-wave order parameter, assume that the Fermi surface
is isotropic in the plane, then, within weak-coupling theory,
$\beta_2/\beta_1=1/2$, and  $K_1=K_2=K_3$ (the later holds up to
particle-hole symmetric terms) \cite{He3,UPt3}, but we shall treat these
coefficients as general parameters.

We shall limit ourselves to solutions which are cylindrically symmetric,
up to an overall gauge transformation.  To this end, it is convenient
to introduce the cylindrical coordinates $(r,\phi)$ for space where
$\phi$ is the angle between $\vec{r}$ and $\hat{x}$, and
define $\eta_{\pm}=\eta_x\pm{}i\eta_y$.  The bulk minimum energy
solutions thus have $\eta_{-}=0$, $\eta_{+}\ne0$, or vice versa.
$\eta_{\pm}$ can be expanded as
$\eta_{+}=\sum_nC^{(n)}_{+}(r)e^{in\phi}$,
$\eta_{-}=\sum_nC^{(n)}_{-}(r)e^{i(n-2)\phi}$, where $n$ is integer and
an extra $-2$ is introduced in the $\eta_{-}$ formula for convenience below.
We have $\vec\eta\cdot\hat{r}=\frac{1}{2}\sum_n(C^{(n)}_{+}(r)+C^{(n)}_{-}(r))e^{i(n-1)\phi}$
and
$\vec\eta\cdot\hat\phi=-i\frac{1}{2}\sum_n(C^{(n)}_{+}(r)-C^{(n)}_{-}(r))e^{i(n-1)\phi}$.
For solutions that are cylindrical symmetric up to a gauge transformation,
only $C_{\pm}^{(n)}$ for one particular $n$ can be finite. Therefore different
solutions are classified by $n$. In these cases,
$C_{\pm}^{(n)}$ can be chosen real without loss in generality.

We shall first assume that the surface at $r=R$ is smooth.
If the order parameter $\vec\eta$ represents the momentum part of
the pairing wavefunction, the component perpendicular to the surface
should vanish \cite{AGD}, thus
\be
  \vec{\eta} \cdot \hat r = 0  \qquad  (r=R),
  \label{b1}
\ee
that is,
\be
  C^{(n)}_{+} + C^{(n)}_{-}  = 0  \qquad (r = R).
  \label{b1a}
\ee
The parallel component $\eta_{\|}\equiv\eta\cdot\hat\phi$
should have vanishing gradient perpendicular to a planar surface \cite{AGD}.
At a surface with a finite curvature, it should satisfy \cite{BF}
\be
  K_1 \frac{\partial}{\partial r} \eta_{\|} = \frac{K_3}{R} \eta_{\|} \qquad (r=R),
  \label{b2}
\ee
hence
\be
  K_1 \frac{\partial}{\partial r} (C^{(n)}_{+} - C^{(n)}_{-}) = \frac{K_3}{R}
  ( C^{(n)}_{+} - C^{(n)}_{-})  \qquad (r=R).
  \label{b2a}
\ee

The differential equations satisfied by $C_{\pm}^{(n)}(r)$ can be found by simple
variation, noting that the total free energy is
$F=2\pi\int_0^Rdrr(\mathcal{F}_b+\mathcal{F}_g)$.
We remark here that the boundary conditions eq.(\ref{b1}-\ref{b2a})
guarantee that there are no net surface terms proportional to
the variation $\delta{}C_{\pm}^{(n)}(R)$ of the order parameters at the surface,
when we integrate by parts the gradient terms.  They also guarantee
that the normal component of the current vanishes for arbitrary
choice of order parameter profiles \cite{BF}, as should be
the case of impenetrable walls at $r=R$.

\begin{figure}
\begin{center}
\rotatebox{0}{\includegraphics*[width=75mm]{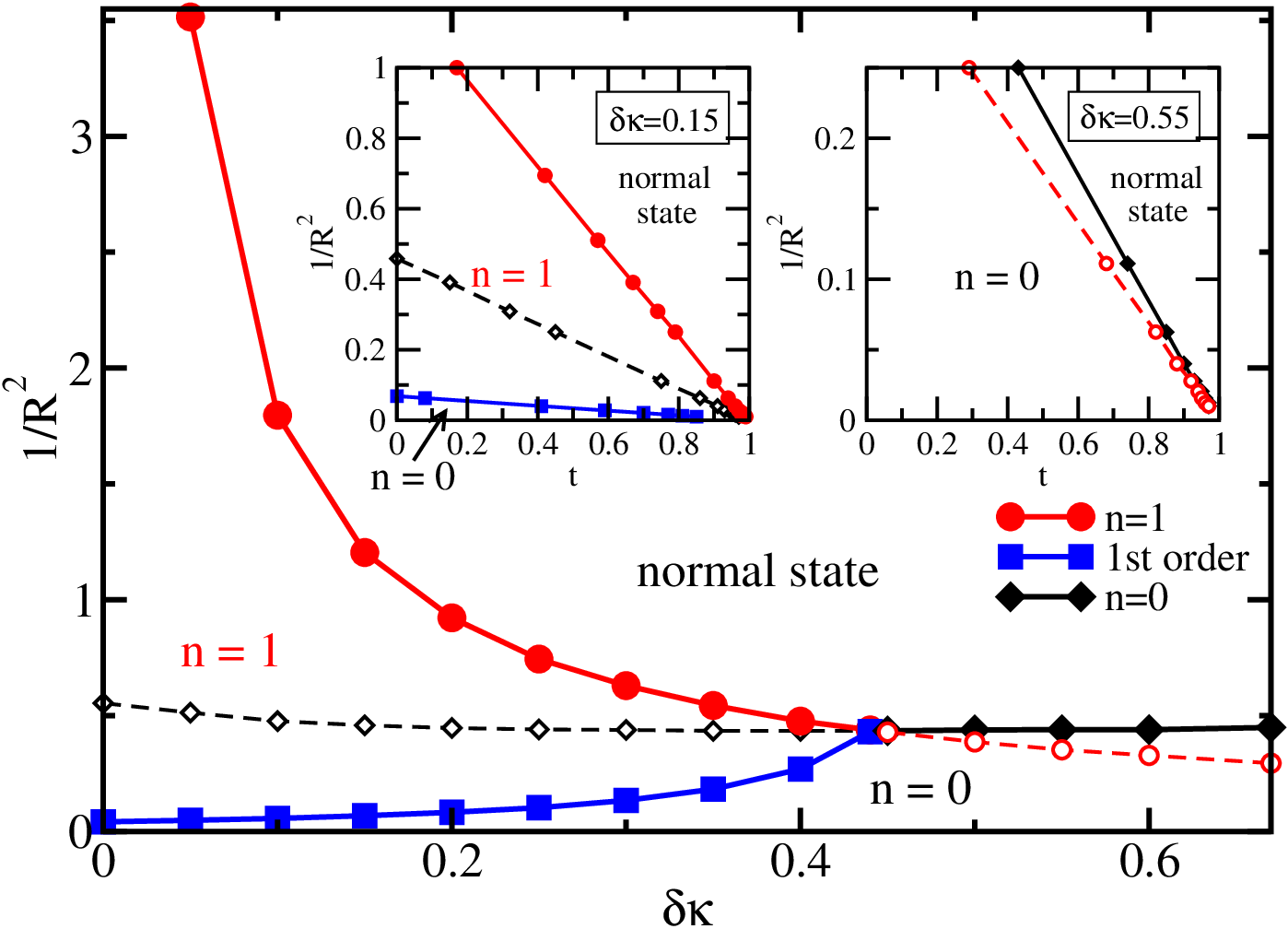}}
\caption{Zero field Phase diagram: $R$ is in unit of
$\xi\equiv\sqrt{K_{123}/\alpha'}$.
The parameters are $\beta\equiv\beta_2/\beta_1=0.25$,
$\kappa_1=1/3+\delta\kappa$, $\kappa_2=\kappa_3=1/3-\delta\kappa/2$,
with $\kappa_i\equiv{}K_i/K_{123}$.
Inset: R-t phase diagrams for different $\delta\kappa$'s.
}\label{fig1}
\end{center}
\end{figure}

First consider zero external magnetic field.
It is worth having some analytic solutions before we show the numerical results.
The transition temperature $t_c^{(1)}$ from the normal to $n=1$ state can be found analytically.
The $n=1$ solutions are time-reversal symmetric, and the free energy
is invariant under $C_{+}^{(1)}\longleftrightarrow\pm{}C_{-}^{(1)}$.
Since the boundary conditions are also symmetric under this transformation,
the equations for $C_s^{(1)}(r)\equiv\frac{1}{2}(C_{+}^{(1)}+C_{-}^{(1)})$ and
$C_d^{(1)}(r)\equiv\frac{1}{2}(C_{+}^{(1)}-C_{-}^{(1)})$
decouple.  One possible solution is $C_s^{(1)}(r)\equiv0$
(hence $\vec{\eta}\cdot\hat{r}=0$ for all $r$), $C_d^{(1)}(r)\ne0$, {\it i.e.}
$C_{+}^{(1)}(r)=-C_{-}^{(1)}(r)=C^{(1)}(r)$. \cite{other} After linearizing in the order parameter, it
can be shown that
$C^{(1)}(r)\propto{}J_1(\sqrt{\frac{\alpha'(1-t_c^{(1)})}{K_1}}r)$, the Bessel function of the first kind.
With eq.(\ref{b2a}),
we find the relation
\be
  (1-t_c^{(1)})=\frac{K_1}{\alpha' R^2} a^2 \ ,
  \label{eq-rt}
\ee
where $a$ should satisfy $aJ'_1(a)=\frac{K_3}{K_1}J_1(a)$.
The critical temperature is suppressed by a factor $\propto1/R^2$.
Therefore, we can find the critical radius via $R_c=a\sqrt{K_1/\alpha'}$, which is the
minimum radius of the system to maintain the superconductivity with $n=1$
at zero temperature. Note that, due to eq.(\ref{eq-rt}), it is more
convenient to set the vertical axes of R-t phase diagram to be $1/R^2$, as
shown in Fig.\ref{fig1}. For this solution, we note that
$a\to0$ as $K_3\to K_1$, which means that in the weak-coupling limit the
transition temperature for our disc with smooth boundary is not at all
affected by the finite radius and in fact is the same as that of the bulk,
hence $1/R_{c}^{2}\to\infty$.
This can also be shown within the quasi-classical approximation, \cite{pf-weak},
and is therefore not an artifact of the GL approximation.
The second case with analytic solution is $K_2=K_3=0$.
The order parameters near the critical point is $C_{+}^{(n)}=c_{n+}J_n(x)$
and $C_{-}^{(n)}=c_{n-}J_{n-2}(x)$. Here $x=r\sqrt{\frac{\alpha'(1-t)}{K_1}}$.
Using the boundary conditions (\ref{b1a}) and (\ref{b2a}),
we can find the R-t relations for all $n$'s.
We find that, in this case, the state with the highest
transition temperature is $n=0$, which is
degenerate with $n=2$.

\begin{figure}
\begin{center}
\rotatebox{0}{\includegraphics*[width=70mm]{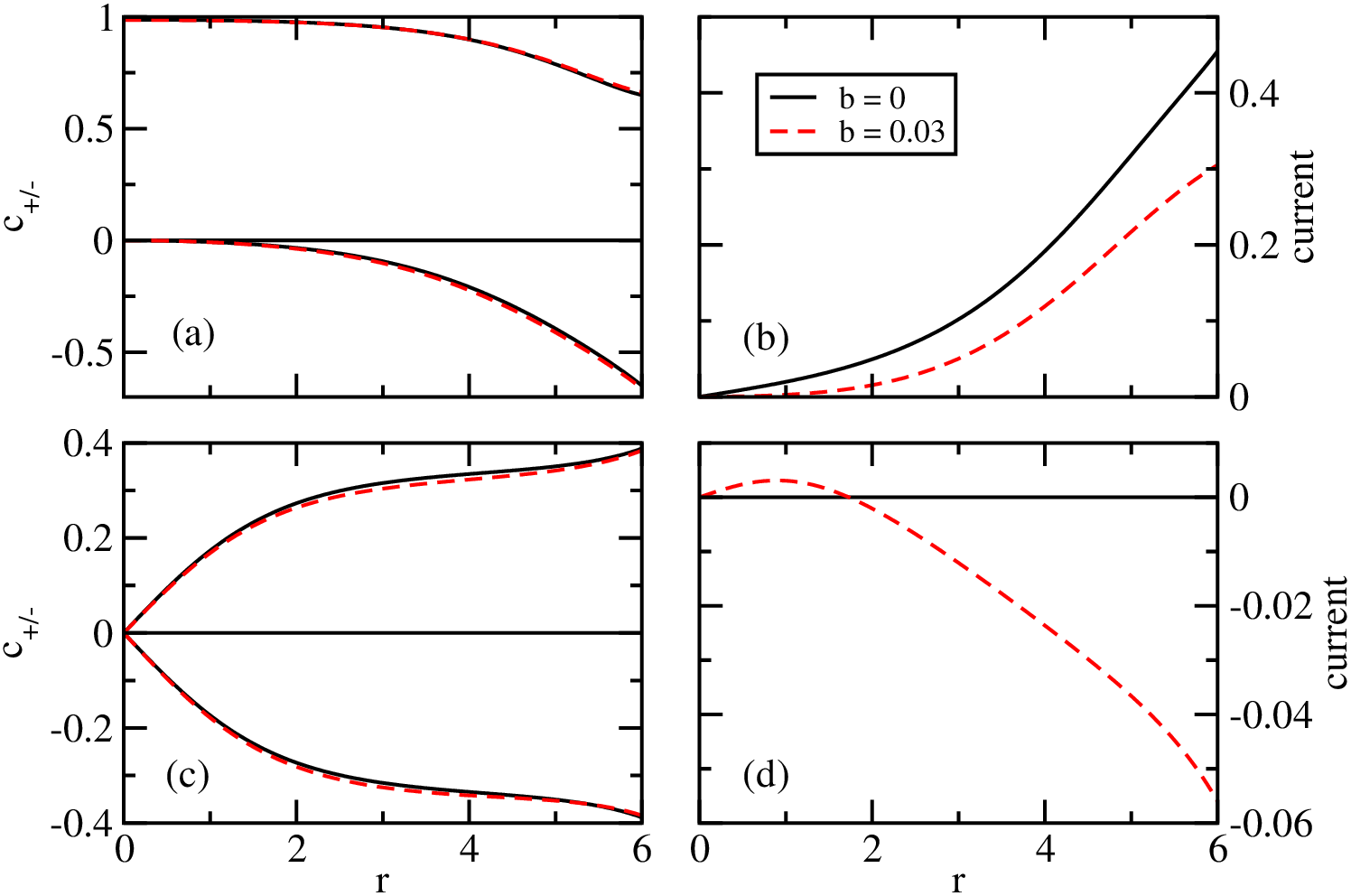}}
\caption{Order parameters (a,c) of the ground states and the
corresponding current distributions (b,d) for zero (black line)
and finite field (red dash). $\delta\kappa=0$,
$\beta=0.25$, $R=6$. The upper
plots are for $n=0$ and $t=0$; the lower plots are for $n=1$ and $t=0.7$.
The order
parameters are normalized to $\sqrt{\alpha'/\beta_1}$,
the external field $B$ is normalized as $b\equiv\frac{2eB}{c}\xi^2$ (roughly to
the bulk upper critical field), $\delta\kappa$ and $\beta$ as defined in
the caption of Fig.1.
}\label{fig2}
\end{center}
\end{figure}

Our obtained phase diagrams are summarized in Fig.\ref{fig1}.
Only $n=1$ and $n=0$
(or equivalently, its time-reversed partner $n=2$ with $C_{\pm}^{(0)}(r)\to{}C_{\mp}^{(2)}(r)$)
can be ground states.
The left inset shows the kind of R-t phase diagram for
$0\leq\delta\kappa\lesssim0.44$ \cite{nk}.
When the radius is smaller than a critical value, $R_c^{(1)}$
(y-intercept of red circle line),
the system does not have any superconducting phase. If the radius is between
$R_c^{(1)}$ and $R_c^{1st}$ (y-intercept of blue square line),
which corresponds to the first order phase
transition from $n=1$ to $n=0$ with lowering temperature, it just has the
superconducting phase with $n=1$ after the second order phase transition (red
circle line). For the radius larger than $R_c^{1st}$, we have second order
phase transition from the normal state and then a first order phase transition (blue
square line) to a spontaneously time-reversal-symmetric-broken superconducting
state ($n=0$).
In the plot, we still show  where
the normal state would have become unstable toward the $n=0$ state
by a dashed line, which does not correspond to a real phase transition
for the system since it is below the $n=1$ transition.
On the other hand, if $\delta\kappa$ is large enough
($\gtrsim0.44$)
, the R-t
phase diagram should be similar to the right inset of Fig.\ref{fig1}. The
system just has the possibility to be in the superconducting state with $n=0$
at low temperature (the unphysical $n=1$ instability line from the normal
state is also shown as dashed). Transition lines
 corresponding to second
order phase transition from the normal state in R-t phase diagrams are linear
within GL theory.
For the first order transition lines,
we found numerically that they are still practically linear.
The main Fig.\ref{fig1} displays the critical radii at zero temperature.
This figure can also be regarded as a plot of the critical radii at finite
temperatures after rescaling of the vertical axis by the factor $(1-t)$.
The linear relations between $1/R^{2}$ and the critical temperatures are artifacts of
GL theory, but we expect that the phase diagrams in Fig.\ref{fig1} are still
qualitatively valid.

In plotting Fig.\ref{fig1}, $\beta=0.25$ was used.  The second
order phase transition lines are independent of this ratio. \cite{nbeta}
For larger (smaller) $\beta$, the first
order transition temperature (and thus the corresponding
$1/R_c^{1st}$) is higher (lower), thus increasing (decreasing) the
stability region for the $n=0$ phase.
 We also note that though our specific calculations are for a circular disc,
the $n=0$ and $n=1$ states are distinguishable by time-reversal symmetry,
and so the phase transition between them should exist even for other
geometries.

To get more understanding of the phase diagram, we consider the
order parameters in some detail.
An example is as shown in Fig.\ref{fig2}.  Here we have chosen $\delta\kappa=0$.
(We shall mostly be considering $0\leq\delta\kappa\lesssim0.44$ for the rest of this paper).
At temperatures below the first order transition temperature $t_c^{1st}$ ($=0.32$ here),
it is a time-reversal-symmetry-broken state
with $n=0$ (or $n=2$).
The flat parts of the order parameters around the center reflects the
characteristics of the bulk system, with $C_{+}\ne0$ but $C_{-}\approx0$.
This is the preferred configuration at low temperatures, or equivalently,
for large samples.
For $t>t_c^{1st}$, the ground state becomes $n=1$. The vortex structure
at the center is shown in Fig.\ref{fig2}(c).
The boundary condition (\ref{b1}) admits only the parallel component
$\vec\eta\cdot\hat\phi$ near the edge of the sample, and,
being at a higher temperature, the radial component
$\vec\eta\cdot\hat{r}$ has not nucleated. This implies that we have $n=1$ \cite{transition},
which is then the preferred configuration at higher temperatures or
intermediate size grains.
The phase diagram reflects the competition between the
boundary effect, which favors $n=1$
(for $\delta\kappa<0.44$),  versus the bulk, which prefers $n=0$ (or $n=2$).

At zero field, the $n=0$ ($n=2$) state has surface current \cite{surfacecurrent}
in $+$ ($-$) $\hat\phi$ direction (Fig.\ref{fig2}(b)), hence a magnetic moment
$\mathcal{M}$ along $+z$ ($-z$).
The $n=1$ state, being time-reversal symmetric, has no surface current
even though there are vortices at the center.

Now we consider an external magnetic field $B$ along the $+$z direction.
First we consider very small fields. The degeneracies between $n=0$
and $n=2$ are lifted.  $n=0$ is favored since its magnetic moment $\mathcal{M}$ is
parallel to the external field.
 The reduction of current near the edge due to external
field in Fig.\ref{fig2}(b) is due to the Meissner effect.
For the time-reversal-symmetric state, with the applied magnetic
field, the system has negative current near the edge and positive current
around the vortex.

\begin{figure}
\begin{center}
\rotatebox{0}{\includegraphics*[width=40mm]{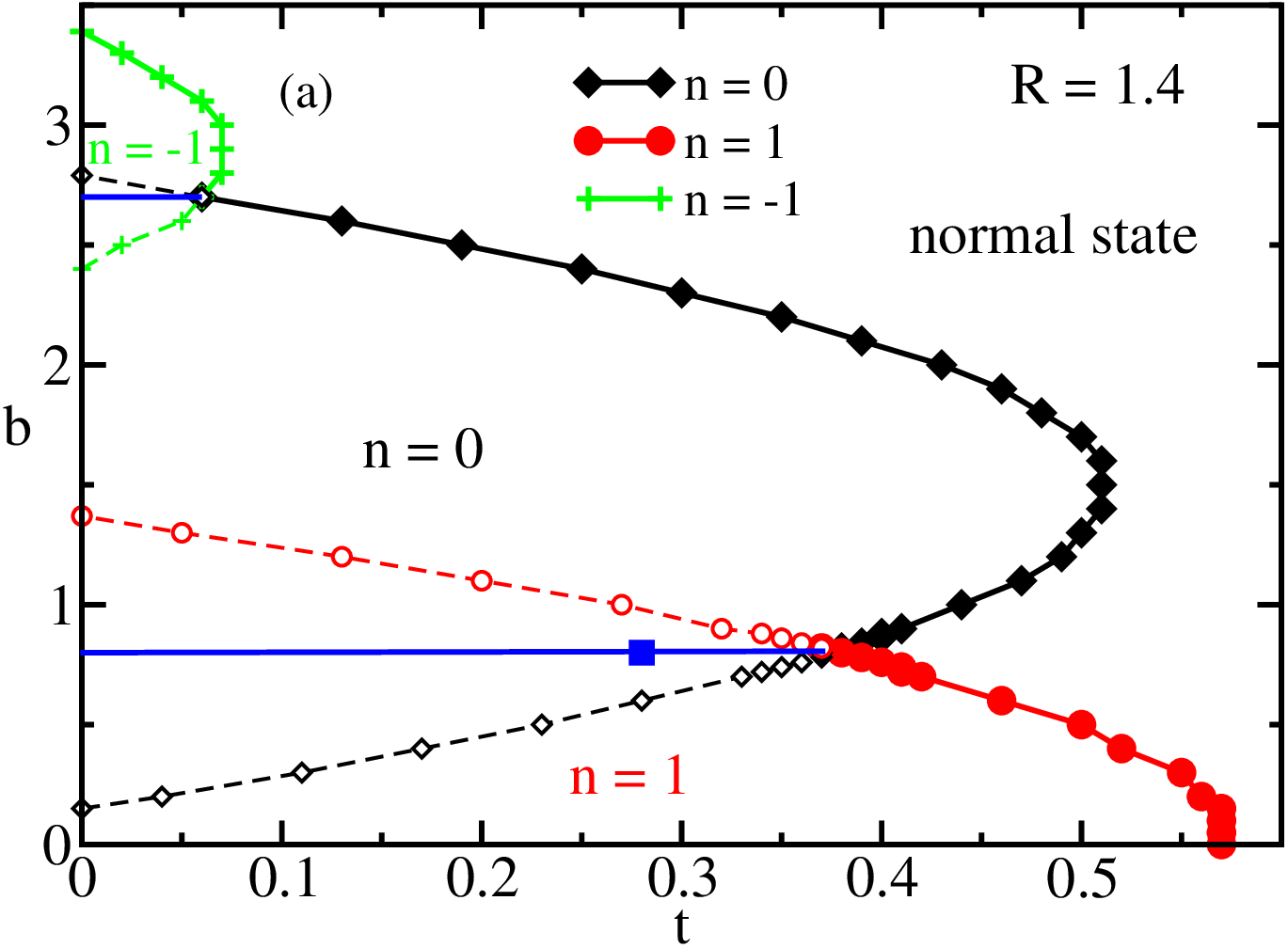}}
\rotatebox{0}{\includegraphics*[width=40.7mm]{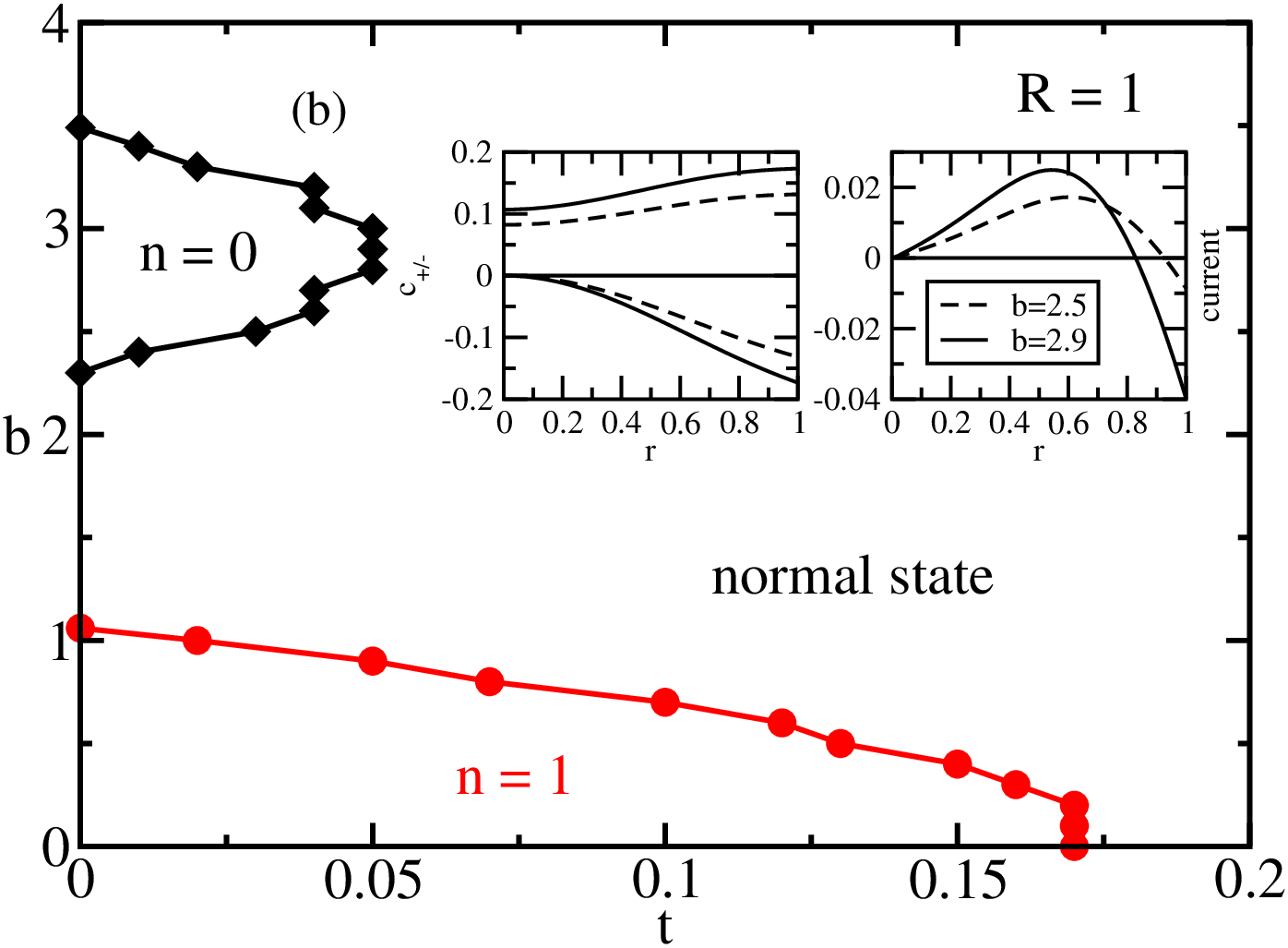}}
\caption{b-t phase diagram: $R=1.4$ and $1$,  $\delta\kappa=0.15$,$\beta=0.25$.
Inset:  order parameter and current at $b=2.5$ and $2.9$, $t=0$.
}\label{fig3}
\end{center}
\end{figure}

For larger fields, the phase diagram can be modified.
In this paper, we focus on discs with
small radii and small magnetic field.
(At larger grains or fields, cylindrical symmetry can be
broken due to the possibility of vortex lattice.
We ignore this possibility in this paper).
An example is shown for $R=1.4$ in Fig.\ref{fig3}(a).
The $n=1$ state is always suppressed by the external field
due to the kinetic energy of the Meissner current. However,
for $n=0$, due to its spontaneous magnetic moment at zero field, it is first {\em enhanced}
by the field, then eventually suppressed at larger fields.
Hence, above a certain
field, $n=0$ can become more favorable than $n=1$.
The transition to the superconducting state can thus become reentrant
as a function of magnetic field.  At still higher fields,
other $n$'s (such as $n=-1$ here) can become the ground state.
For even smaller grains, the reentry
of superconductivity becomes more significant. An example is
shown in Fig.\ref{fig3}(b),
where superconductivity disappears completely for intermediate fields.
This reentrance of superconductivity is similar to the Little-Parks effect
in s-wave superconductors in multi-connected geometries \cite{lp-effect}.
We note that the enhancement or suppression of superconductivity
by $B$ can be understood from the net magnetic moment of the grain,
(hence the current, see inset of Fig.\ref{fig3}(b))
since $\frac{\partial{}F}{\partial{}B}\propto-\mathcal{M}$.

\begin{figure}
\begin{center}
\rotatebox{0}{\includegraphics*[width=36mm]{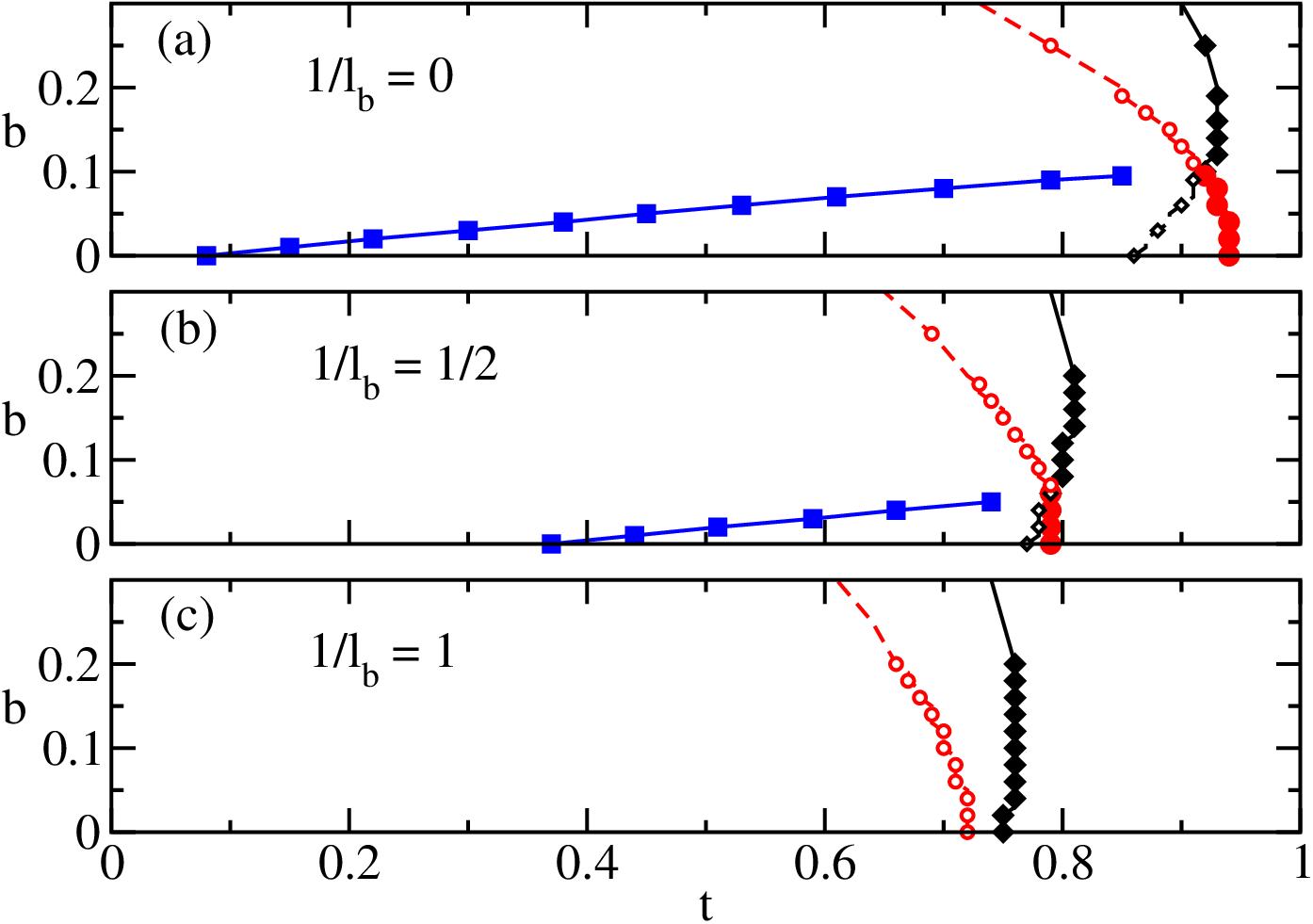}}
\rotatebox{0}{\includegraphics*[width=45mm]{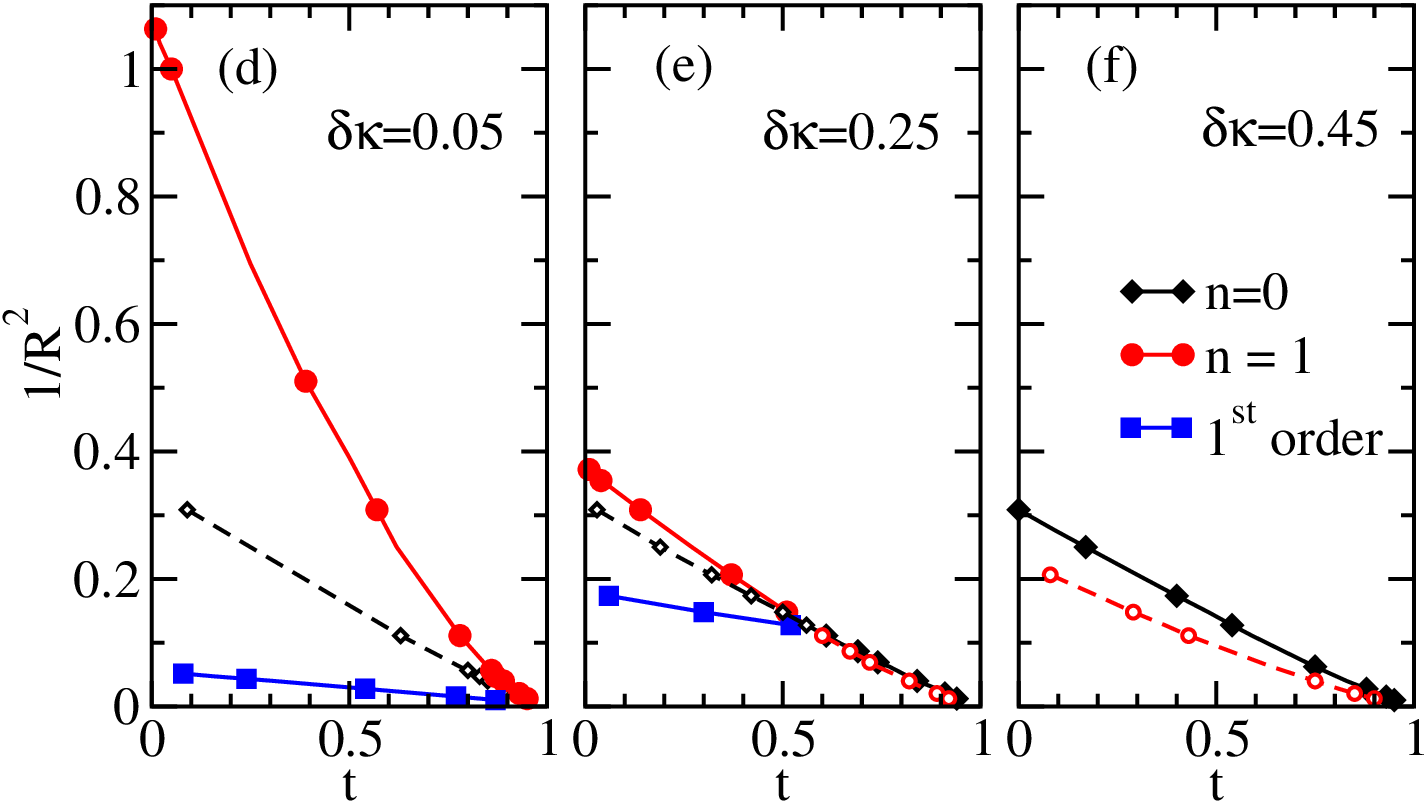}}
\caption{(a-c) Phase diagram: effect of rough boundary on the b-t phase diagram
with $R=4$ and $\delta\kappa=0.15$. (d-f) Zero-field phase diagram with rough
boundary for different $\delta\kappa$'s. $l_b=2.0$.
}\label{fig4}
\end{center}
\end{figure}

Recently, \cite{Huo11} studies a p-wave superconductor in a disc, solving
the Bogoliubov-deGennes equation together with a weak-coupling gap equation,
assuming a cylindrically symmetric Fermi surface and also a smooth boundary at $R$.
Our results for $\delta\kappa=0$ should then be applicable,
but are different in many ways from theirs.
In zero field and some $R$, they showed a transition from the normal state to
the $n=0$ state.  We however found that the transition from the normal state should always be
first to the time-reversal symmetric $n=1$ state, though the system can
make a first order transition later to the $n=0$ state at a lower temperature
for grains that are not too small (Fig.\ref{fig1}).
While both they and we found reentrance of superconductivity as function of magnetic
field, our reentrance is always to a state which is connected with
the one with broken time-reversal symmetry ($n=0$, $-1$, etc here),
but theirs is into a state that is connected with the time-reversal symmetric
one ($n=1$) in zero field (Fig.\ref{fig3}).
Also, we did not find any reentrant behavior as a function of
temperature, in contrast to \cite{Huo11}.

In the above, we have assumed a smooth boundary at $r=R$.  In order to mimic an
imperfect boundary, we introduce the diffusive boundary term,
$\mathcal{F}_s=K_1 C_d^2/l_b$, for the free energy.\cite{Sigrist} Here
$C_d=(C_+-C_-)/2$ and positive $l_b$ is the extrapolation length due to
boundary scattering.
This energy term will modify the boundary condition, eq.(\ref{b2a}). The factor
in front of the order parameter of right hand side becomes $(K_3/R)-(K_1/l_b)$,
as obtained in \cite{AGD} using a more microscopic consideration.
Fig.\ref{fig4}(a-c) show the effect of rough boundary
for $R=4$.
The $T_c$ for $n=1$ is suppressed more than that for $n=0$. However, the $T_c^{1st}$
for the 1st order phase transition becomes larger for rough boundary.
The system can be just in the superconducting state with $n=0$ for sufficiently
rough boundary (as shown in Fig.\ref{fig4}(c)). Fig.\ref{fig4}(d-f)
are the R-t phase diagrams for different $\delta\kappa$'s
without external field.
Comparing these phase diagrams with the insets in
Fig.\ref{fig1}, the second order
transition lines have negative curvatures.
The critical radius for the existence of superconductivity at $\delta\kappa=0$ becomes finite.
The critical
radii at zero temperature gives a phase diagram similar to
Fig.\ref{fig1}, except some shift of the curves to the left.

In conclusion, we study the phase transition of a two-component superconductor in a confined geometry.
We find that, in a large order parameter space, the system would exhibit multiple phase
transitions.
While we have mostly been referring to a p-wave superconducting order parameter,
we expect that many of the qualitative features here would remain so long
as the surface affects the two components of the order parameter differently.
These phase transitions can be detected by, for example, measuring the
density of states via tunneling \cite{meso-s}, in grains of size
of order of coherence length.

This work is supported by the National Science Council of Taiwan
under grant number NSC-98-2112-M-001 -019 -MY3.


\begin{thebibliography}{plain}
\bibitem{other} The other possible choice with $C_d^{(1)}(r)\equiv0$,
$C_s^{(1)}(r)\ne0$ with boundary condition $C_s^{(1)}(r=R)=0$
is found to be unfavorable so long as $K_{2,3}>0$.
\bibitem{nk} For $\delta\kappa<0$, the superconducting transition temperature
would actually be {\em enhanced} by the finite radius. We however have not
pursued this peculiar situation further.
\bibitem{nbeta}  Note that this predicts that for $\beta<0$, $\delta\kappa>0.44$,
the normal state first undergoes an instability into a time-reversal
symmetry breaking state upon lowering of the temperature,  even though
that the bulk only favors a time-reversal symmetric state.
\bibitem{transition} $\vec{\eta}\cdot\hat{r}=0$ for all $r$ requires
$C_{+}^{(n)}(r)=-C_{-}^{(n)}(r)$ for all $r$, which is not possible except $n=1$
since $C_{\pm}^{(n)}(r)$ obey different equations. $C_{+}^{(n)}(r)$ has phase winding
$e^{in\phi}$ but $C_{-}^{(n)}(r)$ has $e^{i(n-2)\phi}$ (see the equations in the paragraph
above eq.(\ref{b1})). $C_{+}^{(1)}(r)=-C_{-}^{(1)}(r)$ is possible due to time
reversal symmetry (see discussion above eq.(\ref{eq-rt})).
\end{thebibliography}
\end{document}


\title{Supplemental Material to "p-wave superconductor in a mesoscopic disc"}

\author{Bor-Luen Huang and S.-K. Yip}

\affiliation{Institute of Physics, Academia Sinica, Taipei, Taiwan}

\maketitle

In the text, we have found that, within GL theory with parameters appropriate
to that of a weak-coupling isotropic p-wave superconductor, the transition
temperature for our circular disc with smooth boundary is not at all affected
by the finite radius $R$ and in fact is the same as that of the bulk.
We shall show in this Supplemental Material that this result is true also within the
quasi-classical approximation, and hence this result is not an artifact of
the GL approximation.

The quasiclassical  green's function ($4 \times 4$) matrix $\hat g (\hat p, \vec r, \epsilon_n)$,
a function of  the momentum direction $\hat p$, position $\vec r$ and Matsubara frequency
$\epsilon_n$, obeys
\be
\left[ i \epsilon_n \tau_3 - \hat \Delta (\hat p, \vec r), \hat g \right]
+ i v_f \hat p \cdot \vec \nabla \hat g = 0
\label{qc1}
\ee
where we adopt the same notations as \cite{SR}.
The normal state green's function is given by
$\hat g = - i \pi {\rm sgn} (\epsilon_n) \tau_3$ and
is diagonal.  $\hat \Delta$ only has matrix
elements that are off-diagonal in particle-hole space.
The transition temperature is obtained by the linearized gap equation, and
hence it is sufficient to solve for the off-diagonal part of $\hat g$ to
first order in $\hat \Delta$.  Assuming spin rotationally invariant pairing
interaction, both these off-diagonal parts of $\hat g$
and $\hat \Delta$ have the same spin structure.  Removing this part,
we then simply obtain equations relating the two scalar functions $f$ and
 $\Delta$ which describe separately the off-diagonal parts of $\hat g$ and $\hat \Delta$.
 Eq (\ref{qc1}) then reads
 \be
 ( 2 i \epsilon_n + i v_f \hat p \cdot \vec \nabla) f
  = 2 i \pi ( {\rm sgn} \epsilon_n ) \Delta
  \label{qcf}
  \ee
  With pairing interaction written as $V_1 \hat p \cdot \hat p'$,
  the linearized gap  equation reads
  \be
  \Delta (\hat p, \vec r) = N(0) T V_1 \sum_{n}
  < ( \hat p \cdot \hat p') f (\hat p', \vec r, \epsilon_n) >
  \label{gapeqn}
  \ee
  where the angular bracket denotes angular average over $\hat p'$ and $N(0)$ is the
  density of states at the Fermi level.

  For the bulk, $\Delta$ is position independent, and hence
  \be
  f (\hat p) = \pi \frac{\Delta (\hat p)}{|\epsilon_n|}
  \label{bulkf}
  \ee
  Eq (\ref{gapeqn}) becomes
  \be
  \Delta(\hat p) =  \pi N(0) T V_1 \sum_{n}
  < ( \hat p \cdot \hat p') \frac{ \Delta (\hat p')}{|\epsilon_n|} >
  \label{bulkgap}
  \ee
  which determines the bulk transition temperature $T_c^0$.

  For the circular disc with a smooth, spin inactive boundary, $\hat g$ obeys
  the boundary condition $\hat g (\hat p_{\rm in}, \vec r, \epsilon_n)
   = \hat g (\hat p_{\rm out}, \vec r, \epsilon_n)$ for $\vec r$ at the surface,
   where $\hat p_{\rm in}$ and $\hat p_{\rm out}$ denote the incoming and outgoing
   momentum directions.  Using the cylindrical coordinates $(r, \phi)$ for
   $\vec r$ and denoting the angle between $\hat p$ and $\hat x$ by $\theta$,
   this boundary condition reads
   \be
   f (\theta, R, \phi, \epsilon_n) = f (\pi + 2 \phi - \theta, R, \phi, \epsilon_n)
   \label{bcf}
   \ee
   In the text, we have seen that the highest $T_c$ corresponds to the case $n = 1$
   and the order parameters at $T_c$ have  the form $C_+^{(1)}(r) = - C_{-}^{(1)}(r) = C r$
   where $C$ is a proportionality constant.  The corresponding $\Delta$ then
   has the form
   $\Delta( \hat p, r, \phi) = \vec \eta \cdot \hat p$
   $= \frac{C}{2} r \left(e ^{i (\phi- \theta)} - e^{ -i (\phi - \theta)} \right)$
   hence
   \be
   \Delta( \hat p, r, \phi) = i C r {\rm sin} (\phi - \theta)
   \ee
   We can verify easily that
   eq (\ref{qcf}) can be solved by
   \be
    f (\theta, R, \phi, \epsilon_n) = \frac{\pi}{|\epsilon_n|} i C r {\rm sin} (\phi - \theta)
    \label{f1}
    \ee
    since $( \vec p \cdot \vec \nabla) f = \left( {\rm cos} (\theta - \phi) \frac{\partial}{\partial r}
      + \frac{ {\rm sin} (\theta - \phi)}{r} \frac{\partial}{\partial \phi} \right) f = 0$.
      Thus we have
      \be
      f (\hat p, \vec r, \epsilon_n) =   \pi \frac{\Delta (\hat p, \vec r)}{|\epsilon_n|}
      \ee
      and with this, the gap equation (\ref{gapeqn})
      becomes
     \be
  \Delta(\hat p, \vec r) =  \pi N(0) T V_1 \sum_{n}
  < ( \hat p \cdot \hat p') \frac{ \Delta (\hat p', \vec r)}{|\epsilon_n|} >
  \ee
  Comparison with the bulk gap equation (\ref{bulkgap}) shows that this is
  satisfied with transition temperature $T_c = T_c^0$ at every position $\vec r$.
  This completes the proof.

  Thus we have verified that the transition temperature is unchanged by the finite
  radius $R$.  Since we now only rely on the quasiclassical approximation, this
  result should hold even for $R$ much smaller than the zero temperature
  coherence length, so long as the product of  the fermi momentum $p_F$ with $R$ is
  much larger than one (and $T_c$ much smaller than fermi energy so that
  the quasiclassical approximation holds).